\documentclass[%
 reprint,
superscriptaddress,
 amsmath,amssymb,
 aps,
prl,
nobalancelastpage
]{revtex4-2}

\usepackage{graphicx}
\usepackage{dcolumn}
\usepackage{bm}
\usepackage{placeins}
\usepackage{float}
\usepackage{hyperref}

\begin{document}


\title{A dual-element, two-dimensional atom array with continuous-mode operation}

\author{Kevin Singh}
\affiliation{%
Intelligence Community Postdoctoral Research Fellowship Program, Pritzker School of Molecular Engineering, University of Chicago, Chicago, IL 60637, USA
}%
\author{Shraddha Anand}%

\affiliation{%
Pritzker School of Molecular Engineering, University of Chicago, Chicago, IL 60637, USA 
}%
\author{Andrew Pocklington}%
\author{Jordan T. Kemp}
\affiliation{%
Department of Physics, University of Chicago, Chicago, IL 60637, USA 
}%
\author{Hannes Bernien}%
\email{bernien@uchicago.edu}
\affiliation{%
Pritzker School of Molecular Engineering, University of Chicago, Chicago, IL 60637, USA 
}%

\date{\today}

\begin{abstract}
Quantum processing architectures that include multiple qubit modalities offer compelling strategies for high-fidelity operations and readout, quantum error correction, and a path for scaling to large system sizes. Such hybrid architectures have been realized for leading platforms, including superconducting circuits and trapped ions. Recently, a new approach for constructing large, coherent quantum processors has emerged based on arrays of individually trapped neutral atoms. However, these demonstrations have been limited to arrays of a single atomic element where the identical nature of the atoms makes crosstalk-free control and non-demolition readout of a large number of atomic qubits challenging. Here we introduce a dual-element atom array with individual control of single rubidium and cesium atoms. We demonstrate their independent placement in arrays with up to 512 trapping sites and observe negligible crosstalk between the two elements. Furthermore, by continuously reloading one atomic element while maintaining an array of the other, we demonstrate a new continuous operation mode for atom arrays without any off-time. Our results enable avenues for ancilla-assisted quantum protocols such as quantum non-demolition measurements and quantum error correction, as well as continuously operating quantum processors and sensors.  
\end{abstract}

\maketitle


\section{\label{sec:level1}Introduction}

\begin{figure*}[ht]
\centering
\includegraphics[width = 0.83\textwidth]{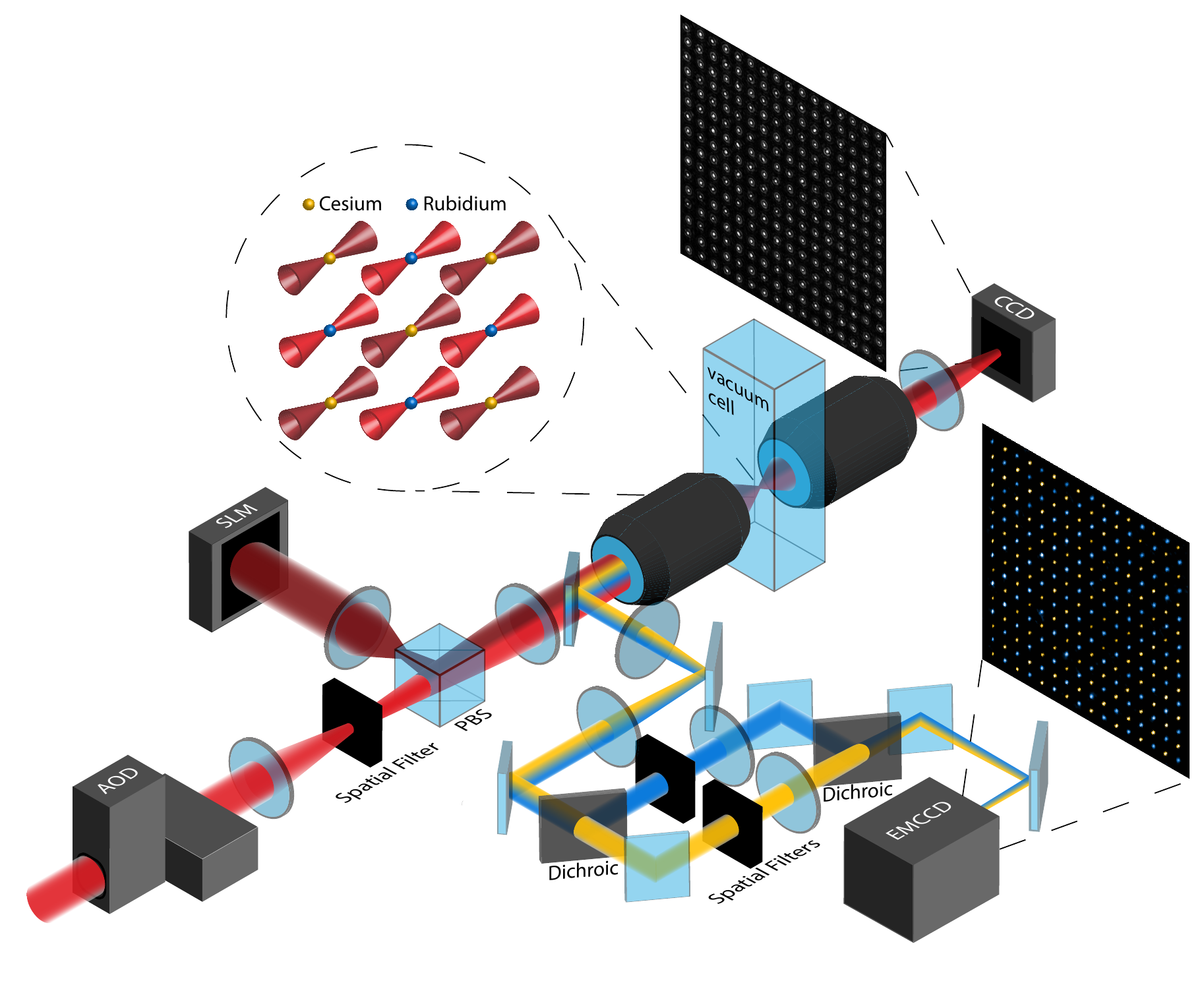}
\caption{\textbf{Trapping and imaging two-dimensional, dual-element arrays of neutral atoms.} Optical tweezers are generated using crossed acousto-optic deflectors (AODs) and a spatial light modulator (SLM) at laser wavelengths of 811 nm (red) and 910 nm (dark red), respectively. An optional spatial filter in the AOD path allows us to mask traps selectively and generate desired geometries. The AOD and SLM trapping arrays are then combined via a polarizing beam splitter (PBS). These combined traps propagate along a shared beam path and are focused by a glass-corrected high numerical aperture microscope objective with NA = 0.65 into a vacuum chamber, thereby creating arbitrary array geometries (shown is a surface-code inspired geometry with two interleaved arrays). An identical objective images the traps onto a charge-coupled-device (CCD) camera to enable feedback-based intensity homogenization. Atomic fluorescence at 780 nm (blue) for Rb and 852 nm (yellow) for Cs is collected using the first objective and reflected along a shared beam path towards an electron-multiplying CCD (EMCCD) camera using a custom dichroic. The signal-to-background is improved by separating the fluorescence wavelengths before the EMCCD and performing individual spatial filtering.}
\label{fig:schematic}
\end{figure*}
\vspace{-10pt}

Realizing large-scale programmable quantum devices with the capability to simulate the behavior of complex processes in physics and chemistry, and to process large amounts of quantum information with high fidelity is at the forefront of science \cite{Feynman1982,Preskill2018quantumcomputingin,PRXQuantum.2.017001,PRXQuantum.2.017003}. A central challenge common to all quantum architectures is how to increase system sizes while maintaining high-fidelity control of and low crosstalk between individual qubits. A universal strategy to address this challenge is to employ a hybrid architecture of multiple qubit modalities, where different types of physical qubits perform distinct functions to evade crosstalk and leverage the advantageous properties of each qubit type \cite{Tan2015,Morton2008}. For instance, Google’s Sycamore quantum processor employs two types of circuit elements made from Josephson junctions for different tasks, with one type used as a set of qubits for processing and the other type used as adjustable couplers to enable low-crosstalk, coherent manipulation of the quantum device \cite{Arute2019}. For quantum dots, the nuclear spins of $^{31}$P donors in silicon have been used as memory qubits with the associated electron spin reserved for processing \cite{Pla2013,Zwanenburg2013}. Analogously, the electron spin of a single nitrogen-vacancy center can be coupled to neighboring nuclear spin qubits ($^{14}$N nuclear spin or $^{13}$C nuclear spins) which act as quantum memories \cite{Bradley2019}. In the ion trapping community, two species of ions are often used, where one species acts as an auxiliary logic qubit to enable sympathetic cooling, state initialization, and detection for a nearby spectroscopy ion \cite{SchmidtP.2005,Bruzewicz2019}. Manipulations and measurements of one species of ion using laser beams have negligible effects on the other ion species because the resonant transition wavelengths have substantial separation \cite{Tan2015}, which can provide, for example, the necessary isolation between memory ions and ions coupled with photonic interfaces needed for the development of scalable ion trap quantum networks \cite{Inlek2017}.

\begin{figure*}[ht!]
\centering
\includegraphics[width = 1.015\textwidth]{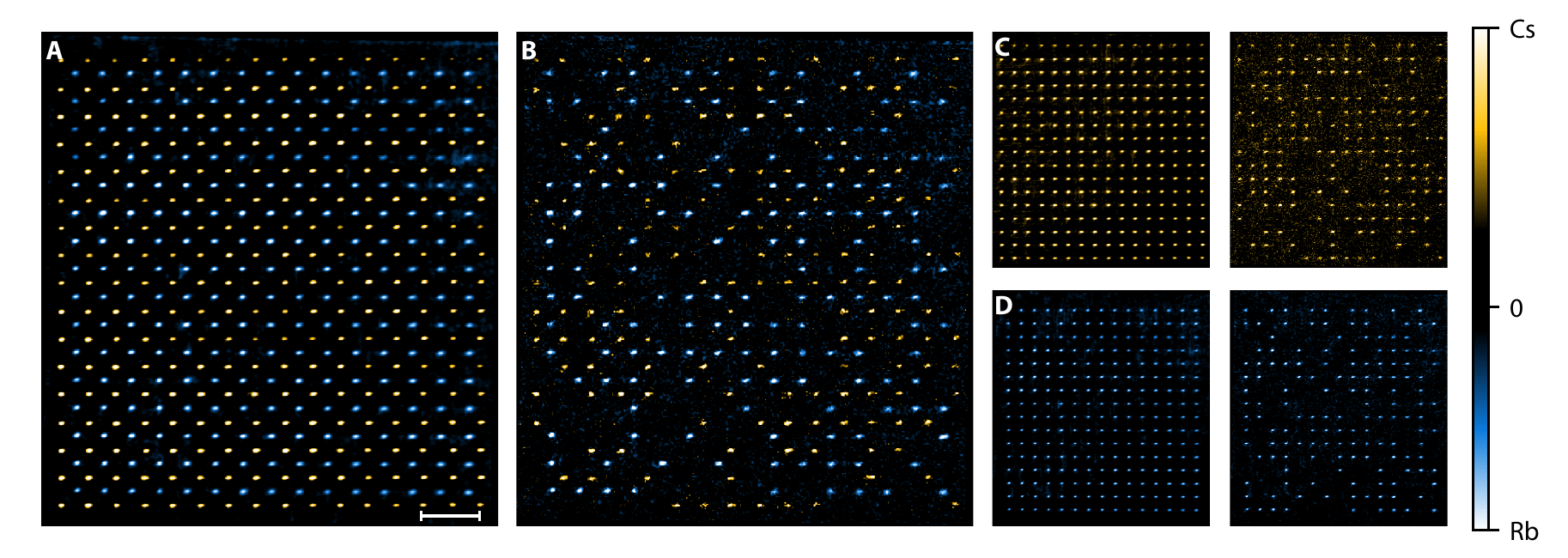}
\caption{\textbf{A dual-element 512-site atom array.} Averaged and single-shot single-atom resolved fluorescence images of the dual-element array with Cs counts in yellow and Rb counts in blue. (A) Averaged fluorescence image of the dual-element array. Scale bar indicates a distance of 20 $\mu$m. (B) Single-shot image of the dual-element array. (C) Averaged and single-shot images of the 17$\times$16 Cs array. (D) Averaged and single-shot images of the 16$\times$15 Rb array. See supplementary text for detailed imaging sequence, parameters, and fluorescence histograms.}
\label{fig:dualarray}
\end{figure*}

Recently, neutral atom arrays have emerged as a promising quantum architecture for pushing the current limits on system sizes \cite{Ebadi2021,Scholl2021}, coherence \cite{Young2020}, and high-fidelity state preparation and control \cite{Madjarov2020,Levine2019,Graham2019,Omran2019,Saffman2016}. In these systems, individual neutral atoms are trapped in arrays of optical tweezers and coherent interactions between atoms are generated by exciting them to Rydberg states. Atom array experiments have reached system sizes of hundreds of atoms \cite{Ebadi2021,Scholl2021,OhldeMello2019}, and recent demonstrations of programmable quantum simulations \cite{semeghini2021probing,de_L_s_leuc_2019,Browaeys2020} and high-fidelity gate operations \cite{Madjarov2020,Levine2019,Graham2019} exemplify the potential of this platform. 

Despite the impressive progress, demonstrations of neutral atom arrays have thus far been limited to single atomic elements, which possess fundamental challenges for readout and control. In particular, the slow and destructive fluorescence-based readout of identical atomic qubits makes it difficult to perform quantum non-demolition (QND) detection, a requirement for quantum error correction, without loss of the qubit state and without nearby atoms absorbing the scattered fluorescence and thereby decohering their quantum states \cite{Morgado2021}. With respect to control, quantum protocols must be halted due to resonant light-scattering and light-assisted atomic collisions when restoring atoms after they have been depleted from the array. These challenges can be overcome by introducing a second atomic element with vastly different transition frequencies into the atom array \cite{Saffman2016}, opening up new hybrid degrees of freedom that can be leveraged to expand and improve control over the quantum system \cite{SchmidtP.2005}. However, neutral atom array architectures with multiple qubit elements have yet to be realized.

Here, we demonstrate a dual-element, two-dimensional atom array constructed from individual rubidium (Rb) and cesium (Cs) atoms trapped in up to 512 optical tweezers. We find that the choice of Rb and Cs atoms enables independent loading, cooling, control, and measurement in the array. This independent control allows us to load Rb and Cs atoms simultaneously into arbitrary two-dimensional array geometries. For instance, we generate arrays where Rb is interleaved within the Cs array in a geometry suitable for surface code operations and stabilizer measurements \cite{Auger2017,Fowler2012}. Moreover, we find that it is possible to load one atomic element into the tweezers while maintaining an array of the other element with no additional losses. This enables the continuous operation of an atomic array without any measurement down-time due to atom loading and initialization, a feature that is inaccessible in single-species atom arrays.

A dual-element array has been a long-sought-after architecture for a myriad of quantum protocols, including quantum sensing assisted by auxiliary qubits \cite{Degen2017}, quantum error-correction \cite{Auger2017}, quantum state manipulation over long time-scales \cite{Saffman2016}, and quantum simulation \cite{Glaetzle2017}. Our results not only open these exciting avenues but also enable the continuous operation of atom array-based quantum processors and sensors.

\section{Results}
\subsection{Dual-element atom array}
We use a dual-wavelength optical tweezer array to load and trap individual atoms from a laser-cooled cloud of Rb and Cs atoms. This optical tweezer array is formed by combining a 2D array of tweezers at 910 nm generated from a spatial-light modulator (SLM) and a separate 2D array of tweezers at 811 nm generated from an acousto-optic deflector (AOD). The wavelengths and laser intensities are chosen such that the 910 nm tweezers and the 811 nm tweezers are element-selective for single Cs and Rb atoms, respectively \cite{Brooks2021}. Control of the phase pattern on the SLM and of the radio-frequency (RF) tones sent to the AOD enables flexible arrangement of the positions of each optical tweezer, allowing us to create arbitrary 2D geometries of Rb and Cs atoms. We detect and resolve individual atoms through a high-NA microscope objective via fluorescence imaging. 

A diagram of our experimental setup is shown in Fig.~\ref{fig:schematic}. The dual-wavelength optical tweezer array is imaged through a secondary microscope objective onto a CCD to confirm the relative alignment of the two 2D tweezer arrays. The focus of the Cs optical tweezers can be brought into the same plane as the Rb optical tweezers by modifying the phase pattern on the SLM. 

As a first demonstration of dual-element loading, we interweave the Rb tweezer array within the Cs tweezer array to form a 512-site dual lattice, where each Rb atom is placed at the center of four Cs atoms on a 2D lattice. After loading the optical tweezer array from a dual-element magneto-optical trap (MOT), we take separate subsequent fluorescence images of the Rb and Cs atoms in the tweezers (for a detailed description of the experimental sequence see supplementary). Averaged and single-shot fluorescence images of the dual-lattice are shown in Fig.~\ref{fig:dualarray}. Figs.~\ref{fig:dualarray}A and~\ref{fig:dualarray}B show example averaged images and single-shot fluorescence images of simultaneously loaded rubidium (blue) and cesium (gold) atoms. Figs.~\ref{fig:dualarray}C and~\ref{fig:dualarray}D show example averaged and single-shot images for only Cs and Rb, respectively. As demonstrated in these images, each atom site is spatially resolved, enabling single-shot single-atom detection of both elements.
 
\begin{figure*}[ht]
\centering
\includegraphics[width = \textwidth]{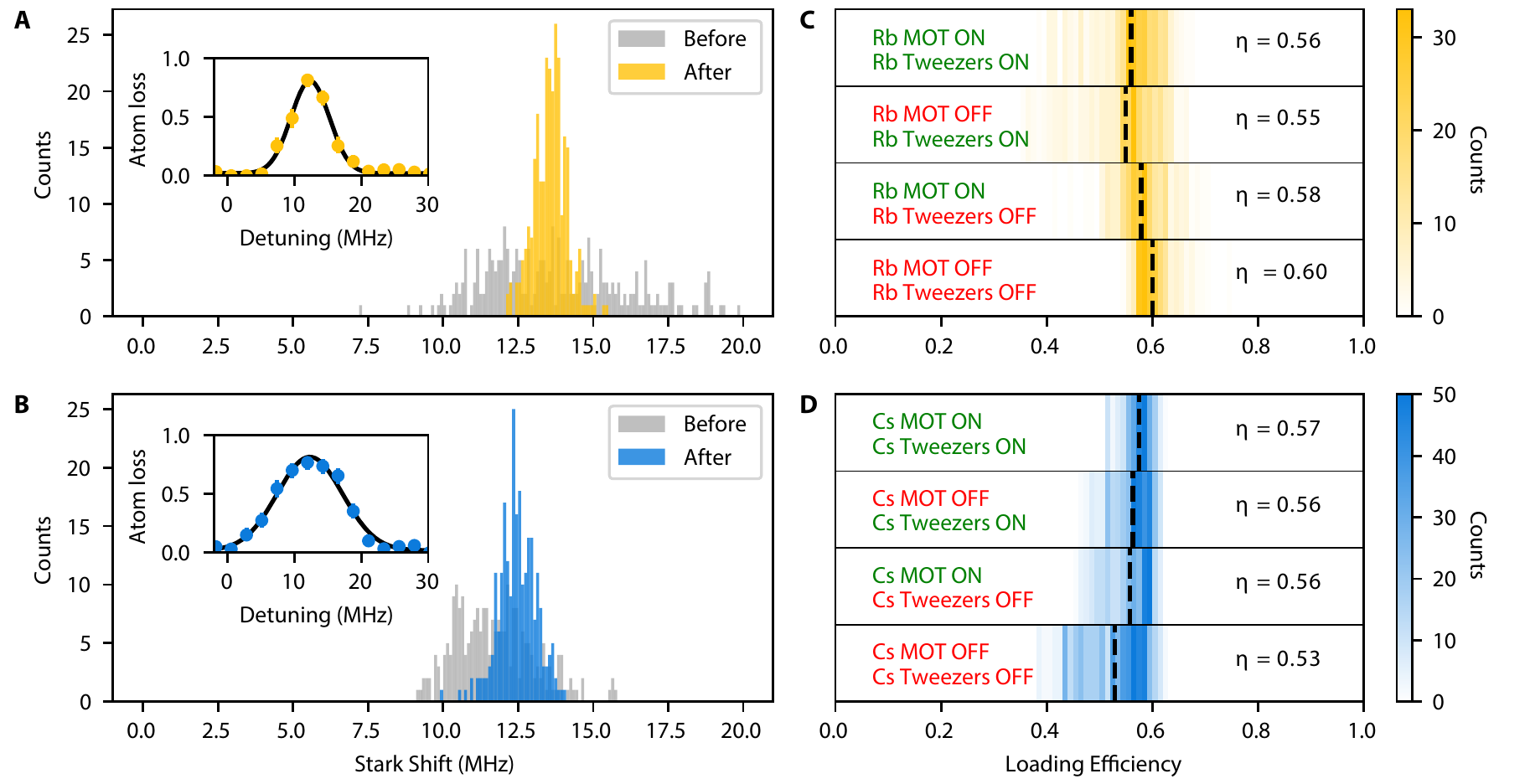}
\caption{\textbf{Homogeneity and loading statistics for the Rb and Cs arrays.} (A) Stark shifts across a 17$\times$16 Cs array before (grey) and after (yellow) trap intensity correction via the weighted Gerchberg-Saxton algorithm based on feedback from the atoms. Inset provides an example Stark shift measurement in which the frequency of a pushout beam is swept to determine the trap-induced AC-Stark shift. (B) Stark shifts across a 16$\times$15 Rb array before (grey) and after (blue) trap intensity correction via the optimization of the RF tones driving the AODs based on feedback from the atoms. Inset provides an example Stark shift measurement. (C) Loading statistics for the Cs array with and without the presence of the Rb MOT and Rb tweezers. Dashed line indicates the average loading efficiency. (D) Loading statistics for the Rb array with and without the presence of the Cs MOT and Cs tweezers. Dashed line indicates the average loading efficiency.}
\label{fig:loading}
\end{figure*}

\vspace{-10pt}
\subsection{Homogeneous arrays and independent loading}
In order to obtain uniform loading across the entire optical tweezer array, it is necessary to homogenize the intensity of the trapping potentials experienced by the atoms. To achieve this, we perform feedback on the amplitude of the RF tones used to generate the Rb tweezers and the phase pattern used to generate the Cs tweezers. As a first step, we use the CCD to homogenize the intensities of each tweezer array to within 2\%. As a second step we directly use the energy shift experienced by the atoms for a more accurate measurement of the tweezer intensities. These energy shifts, called Stark shifts, are shown for Cs (Rb) in the grey histogram in Fig.~\ref{fig:loading}A (\ref{fig:loading}B). For the AOD, we use these measured Stark shifts to weight the amplitude of the RF tones to further homogenize the tweezer intensities. For the SLM, we use the weighted Gerchberg-Saxton algorithm \cite{Kim2019} and replace the target amplitudes with the measured Stark shift values to perform the feedback. The final Stark shifts are shown for Cs (Rb) in the gold (blue) histogram in Fig.~\ref{fig:loading}A (\ref{fig:loading}B) with a uniformity of ~4\% RMS.

We next examine how the loading of the Rb and Cs atoms is affected by the presence of the other atoms’ MOT and tweezer array. In general, one expects inter-species collisional interactions and light-scattering between MOTs of different species \cite{Harris2008,Weber2010}. In our experiment, the large wavelength separation between the laser-cooling transitions at 780 nm (Rb) and 852 nm (Cs) results in a negligible photon-scattering rate for each element with respect to the other element’s laser-cooling light. Additionally, the probability of collisional interactions between the two elements within the tweezers is suppressed because the Cs tweezers are too weak to confine the Rb atoms and the Rb tweezers form anti-trapping potentials for the Cs atoms \cite{Brooks2021}. Figs.~\ref{fig:loading}C and ~\ref{fig:loading}D show histograms of the loading efficiency of each tweezer array with and without the presence of the other atoms’ MOT and with and without the presence of the other atoms’ tweezer array. We find that regardless of the various interaction effects that may occur between the two MOTs and the presence of the dual-array trapping potentials, the average loading efficiency in each tweezer array remains stable and higher than 50\%, in agreement with values measured in single-element arrays operating in the collisional blockade regime \cite{Schlosser2001}.

\begin{figure*}[ht]
\centering
\includegraphics[width = 1.015\textwidth]{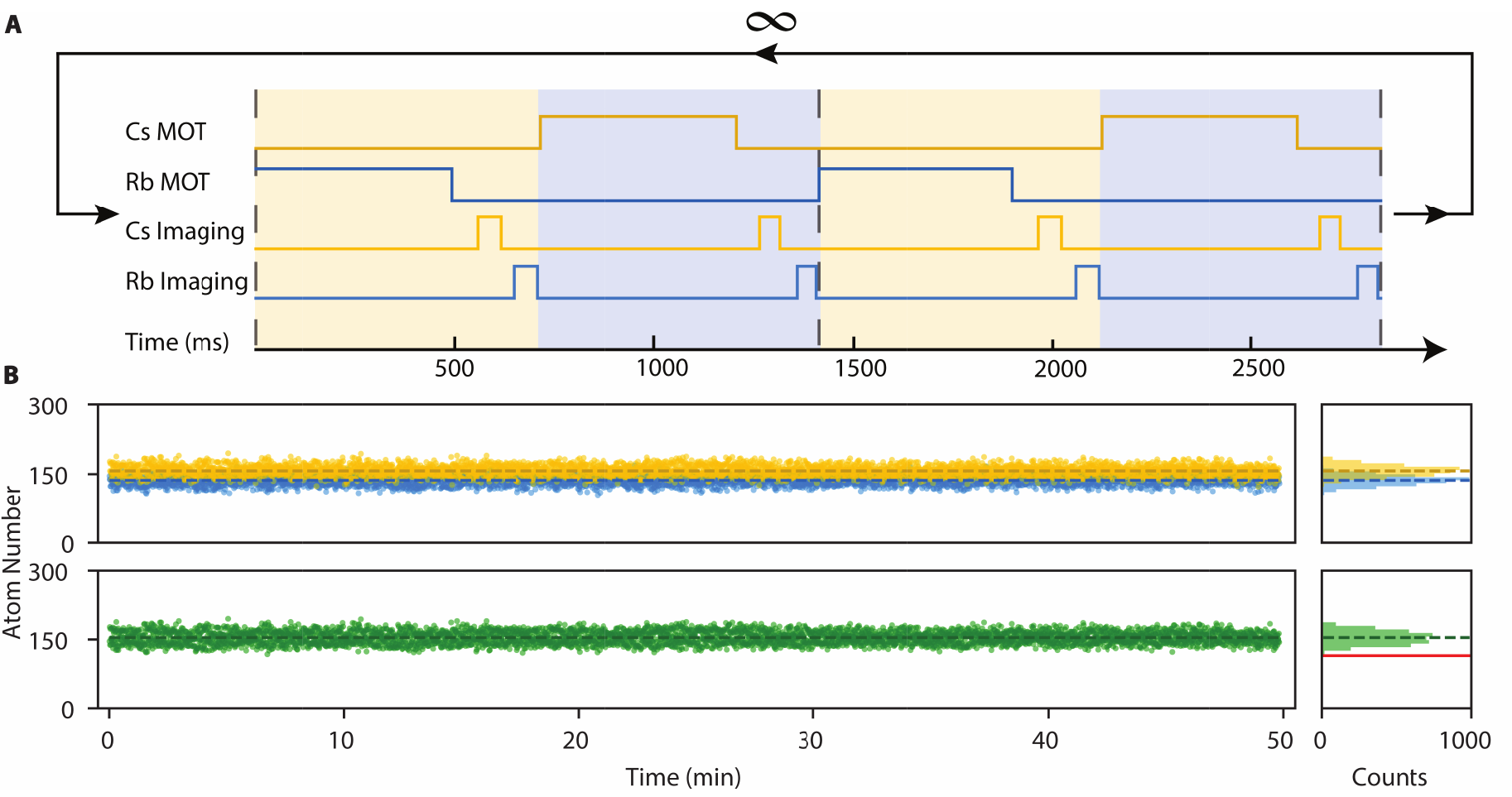}
\caption{\textbf{A continuous-mode atom array.} (A) The pulse sequence used to reload the Rb atoms and Cs atoms into the atom array. Rb (Cs) atoms are reloaded into the array while the Cs (Rb) atoms are held in their optical tweezers. The shaded regions, yellow for Cs and blue for Rb, indicate the atomic element available for manipulation or computation during the specified time window. By performing the pulse sequence repeatedly, a continuously available atomic array can be maintained. (B) The number of Rb (blue) and Cs (yellow) atoms in each image from a 50-minute data run. The dashed lines indicate the average atom counts. The number of atoms available for manipulation as a function of time (green) indicates that the atom array continuously operates with over 115 atoms (red line) at any moment in time.}
\label{fig:perpetuum}
\end{figure*}
\vspace{-10pt}

\subsection{Continuous-mode operation}
The observation that Rb and Cs atoms can be simultaneously loaded into their respective arrays with high efficiency opens up the possibility of loading one of the elements into the tweezer array while holding the other. We investigate this capability with the experimental sequence shown in Fig.~\ref{fig:perpetuum}A, where we continuously alternate which element we load into the optical tweezer array while holding the other. This involves rebuilding a Rb (Cs) MOT while Cs (Rb) atoms are still trapped in the tweezer array. We measure the occupation of each optical tweezer by taking fluorescence images of the Rb and Cs atoms before and after each MOT formation. This procedure allows us to deduce the number of atoms lost due to rebuilding the array of the other atomic element. Remarkably, we observe no additional losses of the held atoms when the other atomic element is loaded in this time period (see methods). 

This independent reloading capability allows us to operate the atom array in a continuous mode, as demonstrated in Fig.~\ref{fig:perpetuum}B, where we repeat the sequence shown in Fig.~\ref{fig:perpetuum}A for 50 minutes. While one element loads into the array, the other element remains idle and available for experiments. By alternating between the elements, we continuously have more than 115 atoms trapped within the tweezer array available for manipulation or computation. We refer to these atoms as data atoms and plot their atom number as a function of time in Fig.~\ref{fig:perpetuum}B bottom. In the context of single-element tweezer arrays, reservoirs of atoms have been used to fill in defects in atom arrays \cite{Manuel2016} or proposed to fill in and re-initialize lost atoms during a computation \cite{Saffman2016}. Operation of the atom array ceases once the reservoir is depleted and only continues once the whole array and the reservoir are reloaded. In our dual-element continuous-mode operation, the newly loaded atoms would not be used to fill in gaps in the array of the other atomic element but rather to continue measurement of a physical quantity or to swap qubit states of the old array into the newly loaded array using Rydberg interaction gates in a manner similar to a “quantum” baton pass. Neither of these applications would be available for single-element atom arrays because of near-resonant light-scattering and light-assisted collisions during the reloading of the MOT. 

\begin{figure*}[ht!]
\centering
\includegraphics[width = \textwidth]{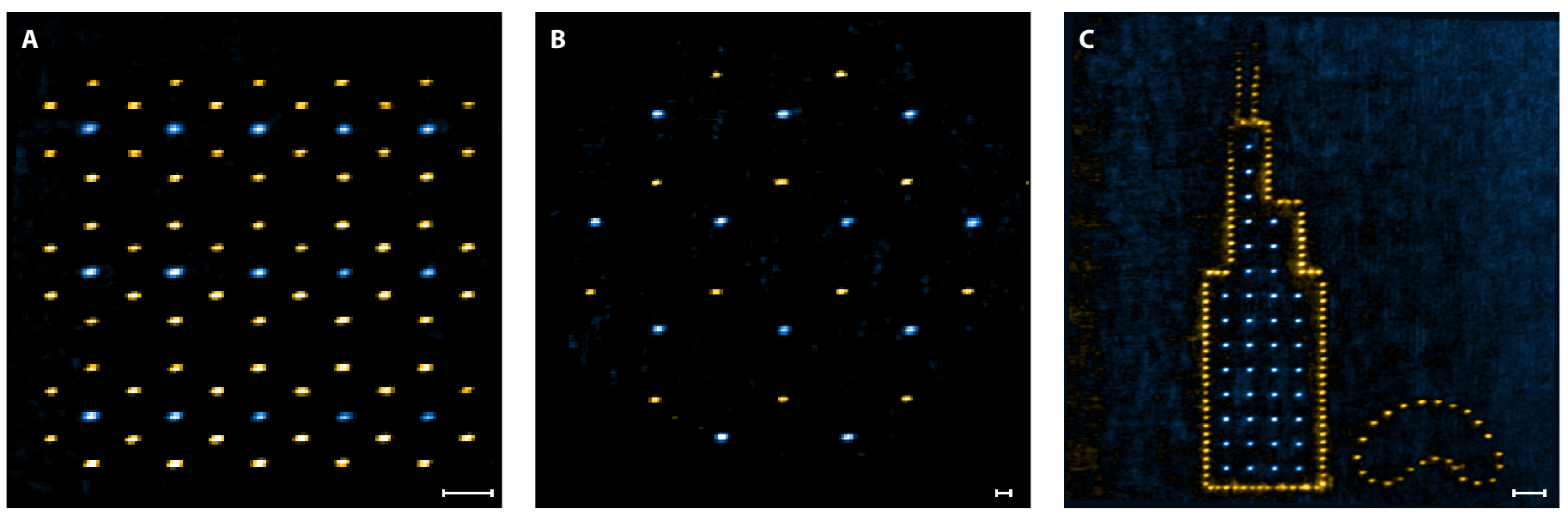}
\caption{\textbf{Arbitrary geometries with dual-element arrays.} Averaged fluorescence images for arbitrary array shapes with Cs in SLM trap sites (yellow) and Rb in AOD trap sites (blue). Scale bars indicate 10 $\mu$m. (A) A Rb-dressed, Cs hexagonal lattice. (B) A bipartite honeycomb lattice. (C) Chicago landmarks: The Sears Tower and The Bean (Cloud Gate). }
\label{fig:geometries}
\end{figure*}
\vspace{-10pt}

\subsection{Arbitrary geometries}
\vspace{-2pt}
To further demonstrate the independent loading and control of the Rb and Cs atoms, we build a variety of dual-element arbitrary arrays shown in Fig.~\ref{fig:geometries}, including a Rb-dressed Cs hexagonal array, a bipartite honeycomb lattice, and two famous Chicago landmarks: the Sears Tower and The Bean (Cloud Gate). While the SLM can directly generate arbitrary trapping arrays, we combine the regularly spaced trapping arrays of the AOD with spatial filtering (shown along the AOD pathway in Fig.~\ref{fig:schematic}) to block specific traps and generate the desired geometries. These results highlight this platform’s capability of placing Rb and Cs atoms in arbitrary geometries with respect to one another, a critical ingredient for engineering qubit interactions and simulating complex models in quantum many-body physics. Additionally, the two arrays can be controllably separated along the out-of-plane dimension by modifying phase pattern on the SLM, opening up research avenues for increasing qubit connectivity, for the study of strongly correlated matter in three spatial dimensions, and for simulating Abelian lattice gauge theories \cite{Notarnicola2020}.  

\newpage
\section{Discussion}
This platform is the first demonstration of dual elements in an atom array experiment and reveals that we retain independent control of the loading, cooling, and imaging of each atomic element. This independent control enables the positioning of single Rb and Cs atoms into arbitrary structures with respect to one another, allowing us to engineer atomic qubit geometries that have important applications in quantum information processing and quantum simulation of complex problems in many-body physics. Additionally, our observation that an atom array can be operated in a continuous mode opens up exciting opportunities in quantum sensing and continuous qubit manipulation. It will be necessary to investigate the coherence of quantum states in one atomic element while the other atomic element is being loaded into the array. Encouragingly, the negligible off-resonant excitation due to the large frequency separation of $2\pi \times$ 32.5 THz and recent results on the coherence in optical tweezers \cite{Guo2020} suggest that coherent manipulation of atomic qubits throughout successive atom loading events is achievable.

Our independent two-element architecture opens up pathways to perform quantum non-demolition measurements and evade crosstalk in neutral atom arrays \cite{Saffman2016}. While this crosstalk can be mitigated using dual-species arrays formed by different isotopes of the same element \cite{sheng2021defectfree}, a dual-element platform benefits from a substantial wavelength separation of atomic resonances \cite{Tan2015,Saffman2016}, species-specific trapping potentials \cite{Brooks2021,Liu2018}, and crosstalk free mutual tunability of homonuclear and heteronuclear Rydberg-Rydberg interactions \cite{Beterov2015} that are important for scaling neutral atom arrays to larger system sizes. With the same atom separations shown in Fig.~\ref{fig:dualarray}, quantum gates using Rydberg interactions can be used to entangle the qubit states from one element serving as a ‘data’ qubit with another element serving as an ‘auxiliary’ qubit, which can then be detected without added perturbations of the ‘data’ qubits. This Rydberg gate can also be used to entangle a single ‘auxiliary’ qubit with a large number of ‘data’ qubits in a single step \cite{Mueller2009}. For these applications, it will be necessary to deterministically load the atoms without defects using standard rearrangement techniques \cite{Ebadi2021,Manuel2016,Daniel2016,Kim2016}. Due to the geometry of our 512 site-dual lattice, simultaneous row and column rearrangements of the Rb and Cs atoms naturally avoid collisions with one another, thereby enabling efficient rearrangement movements. We plan to implement these rearrangement protocols using two SLMs to generate permanent optical tweezers for each element and one AOD to perform simultaneous rearrangement of both elements. Moreover, system sizes can be increased with additional laser power while remaining within the 300-micron field-of-view of our microscope objective. 

With respect to interactions, Rydberg-excitation lasers can be used to either uniformly illuminate the entire array from the side of the glass cell to generate long-range interactions or, with an addition of an SLM, perform site-specific entangling gate operations through the second microscope objective shown in Fig.~\ref{fig:schematic}. Furthermore, one can use Förster resonances between the Rb-Rb, Cs-Cs, and Rb-Cs atoms to tune the strength of interactions between any pair of atoms to be weak or strong with respect to one another \cite{Beterov2015}. The wide tunability of these asymmetric Rydberg interaction strengths enables the exploration of new methods of large-scale multi-qubit manipulation and control, allowing, for example, interactions between one species of atoms to be mediated by the other. Accordingly, several proposals suggest that dual-element architectures using Rb and Cs qubits are well-suited for developing a neutral atom-based coherent quantum annealer \cite{Glaetzle2017} and for fault-tolerant quantum computation with Rydberg atoms \cite{Morgado2021}. These dual-element features make our platform an excellent starting point for quantum sensing assisted by auxiliary qubits \cite{Degen2017} and quantum error correction in neutral atom arrays \cite{Auger2017}.

\section{Materials and Methods}

\subsection{2D and 3D MOTs}
The Rb and the Cs atoms are generated from two alkali metal dispensers placed inside a glass cell (ColdQuanta) and first cooled in a retro-reflected bichromatic 2D magneto-optical trap (MOT) operated at both 780 and 852 nm. A bichromatic push-beam then aids in the transfer of the atoms through a pinhole into a separate vacuum glass cell (JapanCell), where a dual-element 3D MOT traps and further cools the atoms. An ion pump (NEXTorr D500-5) is used to generate ultra-high vacuum in the glass cell with a measured background pressure of $< 10^{-11}$ Torr. 

The MOT beams for both elements share the same beam paths and are generated by two DBR laser modules (Vescent Photonics) at 780 nm (Rb) and 852 nm (Cs). For the Rb (Cs) 3D MOTs, the cycler beams are 12.9 MHz red-detuned from the free space F = 2 → F’ = 3 (F = 4 → F’ = 5) D2 transition and the repump beams are nearly resonant with the free space F = 1 → F’ = 2 (F = 3 → F’ = 4) D2 transition. For both elements, the MOT beam cycler powers are set to the saturation intensities for the relevant transitions with the associated repump power at 10\% of the corresponding cycler power. Both atomic elements are loaded into the optical tweezers with the 3D MOT field gradient set to $\sim$18 G/cm. The 3D MOT beam sizes are irised down to a $\sim$2 mm diameter to minimize stray reflections from the vacuum chamber during imaging. 

\subsection{Dual-element 2D optical tweezer arrays}
The trapping light for Rb and Cs are generated separately by two Ti:Sapphire lasers (MSquared) set to 811 nm and 910 nm, respectively. The optical tweezer array for the Rb atoms is generated by passing 811 nm light through a pair of crossed acousto-optic deflectors (AA Opto Electronic) controlled with RF tones generated by an arbitrary waveform generator (Spectrum). We use an SLM (Holoeye) to imprint a computer-generated hologram on 910 nm laser light to generate the tweezer array for the Cs atoms. A high numerical aperture microscope objective (Special Optics) with NA = 0.65 is used to tightly focus the tweezers down to Gaussian waists of $\sim$0.8 microns within the spatial region of the 3D MOTs.

To homogenize the trap depths, we perform feedback on the intensities measured by a CCD camera and on the Stark shift measurements on the atoms. For the Rb tweezers, this feedback is done on the amplitude of the RF tones sent to the AOD [36]. For the Cs tweezers, the feedback is performed on the target amplitudes in the weighted Gerchberg-Saxton algorithm used to generate the Cs tweezers. Here, we also correct for optical aberrations by scanning and correcting for low-order Zernike polynomials to maximize the measured intensity in the center of the tweezers \cite{Ebadi2021}. 

\subsection{Imaging}
The atoms are detected by taking subsequent fluorescence images of the trapped Cs and Rb atoms at 852 nm and 780 nm, respectively. Fluorescence is separated from the trapping light by a multi-edge dichroic (Laser Zentrum Hannover e.V.) and is collected for a period of 40 ms for each image on an EMCCD (Andor IXON 888) camera. We remove the scattered background light in the images by separating the two imaging wavelengths using a dichroic and performing spatial filtering in the back focal plane of the microscope objective. Additionally, to reduce the background due to the presence of atoms in shallow out-of-plane SLM traps, we use a low-intensity blow-out pulse to remove these weakly bound atoms immediately after loading the tweezer array.  

\subsection{Losses during continuous-mode operation}
For the experimental sequence in Fig.~\ref{fig:perpetuum}A, the average loss rates between successive images, with the MOT reload of the other element occurring between images, were measured to be 0.095 $\pm$ 0.013 for Rb and 0.104 $\pm$ 0.032 for Cs. For the same experimental sequence, we turn off the 2D MOT and 3D MOT for one element to set the baseline loss rate of the other atom. We measure that the baseline Rb loss rate without the presence of the Cs atoms is 0.093 $\pm$ 0.020 and the baseline Cs loss rate without the presence of the Rb atoms is 0.109 $\pm$ 0.032.

\subsection{Statistical Analysis}
The presence of individual atoms in each image is determined by fitting bi-modal distributions to the total fluorescence counts in the vicinity of each tweezer site and setting a threshold between the ‘dark’ and the ‘signal’ modes. Using this information, we extract all relevant statistical quantities such as site-wise loading efficiencies and losses. All error bars presented in this analysis are the Clopper–Pearson intervals for that parameter.

\begin{acknowledgments}
\section{Acknowledgments}
We thank Shankar Menon, Noah Glachman, Jonathan Simon, and Kang-Kuen Ni for fruitful discussions and critical reading of the manuscript. We thank Samantha Lapp and Kin Fung Ngan for their contributions during the initial design and setup of the experiment. We acknowledge funding from the Office for Naval Research (N00014-20-1-2510), the Air Force Office of Scientific Research (FA9550-21-1-0209), the NSF QLCI for Hybrid Quantum Architectures and Networks (NSF award 2016136), and the Sloan Foundation. This research was supported by an appointment to the Intelligence Community Postdoctoral Research Fellowship Program at the Pritzker School of Molecular Engineering administered by Oak Ridge Institute for Science and Education through an interagency agreement between the U.S. Department of Energy and the Office of the Director of National Intelligence.
\end{acknowledgments}


\bibliography{a2d2}
\bibliographystyle{apsrev4-1}
\onecolumngrid

\clearpage
\section*{SUPPLEMENTARY MATERIAL}
\addcontentsline{toc}{section}{SUPPLEMENTARY MATERIAL}
\setcounter{subsection}{0}

\subsection{Simultaneous loading of rubidium and cesium}
The experimental sequence for loading atoms into the optical tweezer array is shown in Fig.~\ref{fig:sequence}. The dual-wavelength optical tweezer array (811 and 910 nm laser light) remains on during the duration of the experiment. Laser cooling of thermal $^{87}$Rb and $^{133}$Cs atoms begins by turning on the 2D and 3D MOT laser light for $\sim$300 milliseconds. After loading the atoms in the MOT, the magnetic field gradient is extinguished, and the atoms are cooled below the Doppler temperature limit via polarization-gradient cooling (PGC) in 20 ms by lowering the MOT laser intensities and detunings. The laser cooling light is then turned off for 10 milliseconds to allow atoms not trapped in the optical tweezer array to disperse. We find that the SLM-generated tweezer array also includes spurious, out-of-plane traps. We remove any Cs atoms that may be weakly trapped at these sites by applying a weak, nearly-resonant blowout pulse at 852 nm. Two sets of fluorescence images of the Rb and Cs atoms are then taken to measure loading statistics and atom losses.

\setcounter{figure}{0}
\makeatletter 
\renewcommand{\thefigure}{S\@arabic\c@figure}
\makeatother

\newcounter{SIfig}
\renewcommand{\theSIfig}{S\arabic{SIfig}}

\vspace{-20pt}
\begin{figure*}[h!]
\centering
\includegraphics[width = 1\textwidth]{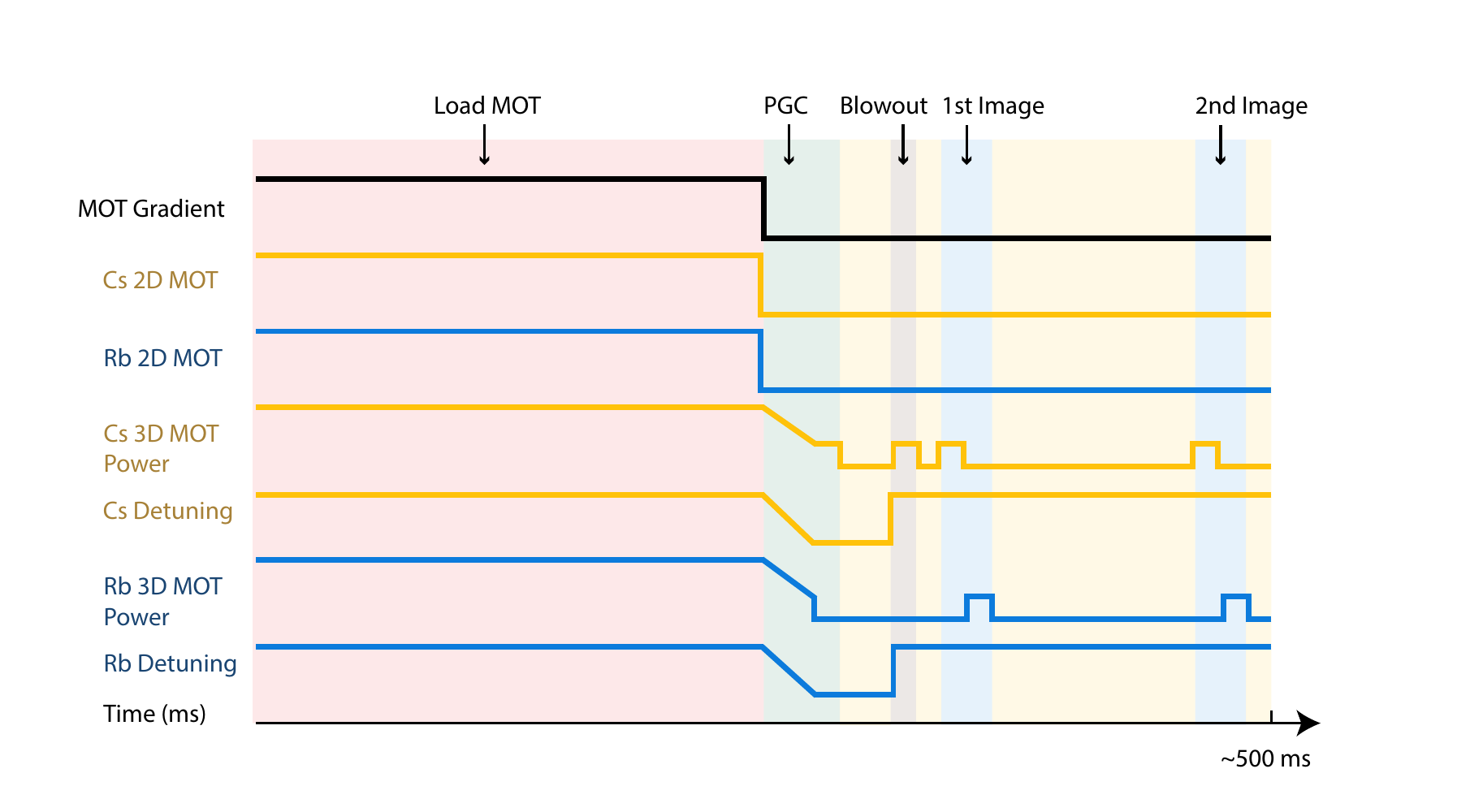}
\vspace{-20pt}
\caption{\textbf{Experimental sequence.} Diagram of the experimental sequence. Cs (Rb) 2D MOT refers to the state (on or off) of the Cs (Rb) 2D MOT cooling beams. We first load a dual-element MOT for $\sim$300 milliseconds. After MOT formation, the 2D MOT beams are turned off and the 3D MOT beam powers and detunings are ramped to perform polarization-gradient cooling (PGC). After a short wait time to allow atoms not trapped in the tweezers to disperse, a Cs blowout pulse is performed to remove Cs atoms trapped in weak out-of-plane SLM traps. Two sets of images are performed to measure atom statistics. The dual-wavelength optical tweezer array at 811 and 910 nm remains on through this entire sequence.}
\refstepcounter{SIfig}\label{fig:sequence}
\end{figure*}
\vspace{-20pt}

\subsection{Optical tweezer and atom characteristics}
Both the 811 nm tweezer array and the 910 nm tweezer array are focused onto the atoms using the diffraction-limited 0.65 NA microscope objective. After passing through this objective, each individual optical tweezer has $\approx 1$ mW of optical power. Using release and recapture measurement of the atoms \cite{Sortais2007}, the Rb atoms are measured to have radial trap frequencies of $\omega_r = 2\pi \times 100$ kHz in the 811 nm array, and the Cs atoms are measured to have radial trap frequencies of $\omega_r = 2\pi \times 60$ kHz in the 910 nm array. Via comparison with Monte Carlo simulations, we measure the temperatures of the Rb atoms in the optical tweezers to be 50 $\mu$K and the temperatures of the Cs atoms to be 30 $\mu$K at our given optical tweezer intensities. These temperatures can be lowered to a few microkelvin via adiabatic cooling by lowering the depth of the trapping potentials \cite{Tuchendler2008}. \\

\begin{figure*}[h!]
\centering
\includegraphics[width = 0.98\textwidth]{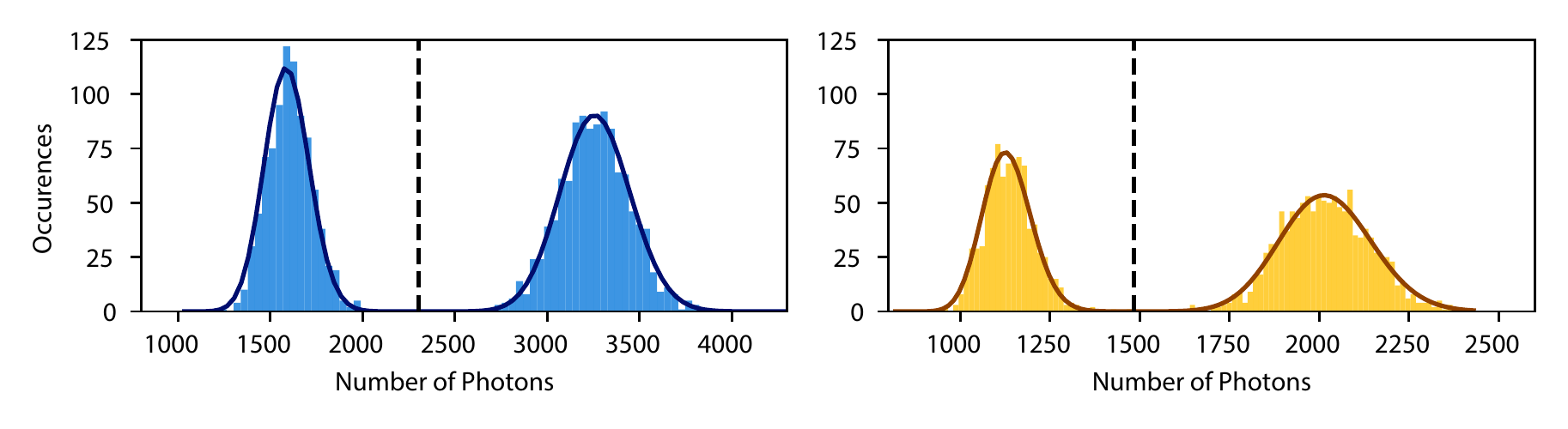}
\vspace{-10pt}
\caption{\textbf{Example atomic fluorescence histograms.} Sample fluorescence counts for a single Rb atom (left) and a single Cs atom (right) collected during a 40 ms imaging time. Thresholds (dashed lines) are placed between the atom and background signals by fitting the fluorescence counts to a bi-modal Poisson distribution and extracting the minimum between the modes.}
\refstepcounter{SIfig}\label{fig:hists}
\end{figure*}

We image the atoms held within the optical tweezers by turning on the 3D MOT beams and collecting the scattered photons from each atom with our microscope objective. These photons are then imaged onto an EMCCD to perform single-site detection of each atom. Example histograms indicating the number of photons collected by a single Rb atom (blue) and a single Cs atom (gold) are shown in Fig.~\ref{fig:hists}. In each histogram, the left peak indicates the number of photons collected when an atom is not present, and the right peak indicates the number of photons collected when an atom is present. The presence of atoms is calculated by fitting these histograms to bi-modal distributions and placing a threshold between the peaks. We never observe Rb atoms in the location of the Cs tweezers (and vice versa).  \\

\begin{figure*}[h!]
\centering
\includegraphics[width = 0.63\textwidth]{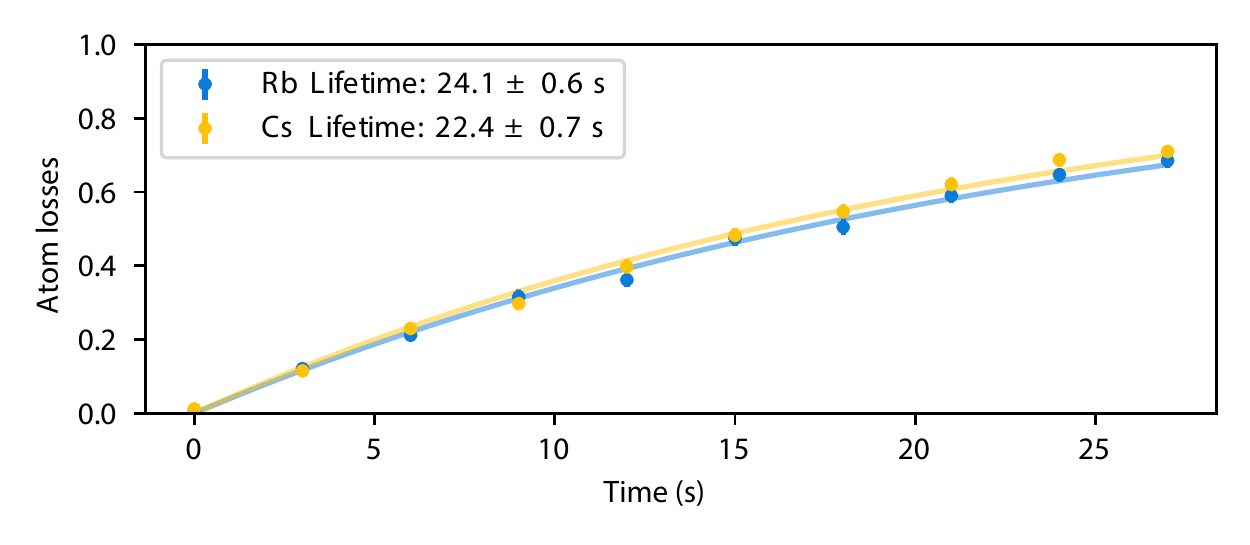}
\vspace{-15pt}
\caption{\textbf{Atom lifetimes in optical tweezers.} Typical atom lifetimes for Rb and Cs atoms under continuous laser cooling with the PGC light while trapped in the 811 and 910 tweezers, respectively. Solid lines indicate exponential fits to the data. Error bars are smaller than the size of the markers.}
\refstepcounter{SIfig}\label{fig:lifetime}
\end{figure*}

Typical lifetimes of the trapped atoms with continuous laser cooling using the PGC light are shown in Fig.~\ref{fig:lifetime}. This lifetime is set by background gas collisions in the vacuum chamber and can be improved in the future with higher vacuum \cite{Covey2019} or with a cryogenic environment \cite{Schymik2021}.

\end{document}